\documentclass[prd, preprint, showpacs, showkeys]{revtex4}
\begin{document}
\date{\today}
\title{Alternative variables for the dynamics of general
relativity}
\author{R. Rosas-Rodr{\'\i}guez}
\email{rrosas@sirio.ifuap.buap.mx}
\affiliation{Instituto de
F{\'\i}sica, Universidad Aut\'onoma de Puebla, Apdo. Postal J-48,
72570, Puebla, Pue., M\'exico}

\date{\today}
\begin{abstract}
A new form of the dynamical equations of vacuum general relativity
is proposed. This form involves the canonical Hamiltonian
structure but non canonical variables. The new field variables are
the electric field $E^{a}{}_{i}$  and the magnetic field
$B^{a}{}_{i}$ which emerge from the Ashtekar's representation for
canonical gravity. The constraint algebra is studied in terms of
the new field variables and the counting of the degrees of freedom
is done. The quantization is briefly outlined.
\end{abstract}

\maketitle
\section{Introduction}
In 1986, the canonical approach to general relativity received new
life by the introduction by Abhay Ashtekar of a new formulation
\cite{Ash1}. (See also \cite{Ash2}.) In this formulation one can
use a (complex) $SO(3)$ spatial connection as coordinate for the
gravitational phase space instead of the 3-metric introduced by
Arnowitt, Deser and Misner (ADM) \cite{ADM}. Ashtekar variables
led to a considerable simplification of the constraints associated
with the Hamiltonian formulation of Einstein's theory. Indeed,
Ashtekar's constraints are polynomials in the canonical variables.
Ashtekar's canonical gravity allows some progress in the direction
of a quantum theory of gravity.

On the other hand a common framework has emerged which extends the
structure of Hamiltonian mechanics to infinite-dimensional
systems. The Hamiltonian formulation is usually obtained from the
Lagrangian formulation by means of the Legendre transformation,
but in the case of fields this canonical procedure presents
difficulties since not always the momentum densities are
independent of the field variables, which is usually mended by the
introduction of constraints. Nevertheless, it is possible to avoid
these complications and give a Hamiltonian formulation for a given
continuous system, without making reference to the Lagrangian
formulation, if its evolution equations can be written in the form
\begin{equation}
\dot{\phi} _{\alpha}= D_{\alpha\beta} \frac{\delta
H}{\delta\phi_{\beta}}, \label{eq1}
\end{equation}
where the field variables $\phi _{\alpha}$ $(\alpha=1,2,...,n)$
represent the state of the system, $H$ is a suitable functional of
the $\phi _{\alpha}$, $\delta H/\delta\phi _{\beta}$ is the
functional derivative of $H$ with respect to $\phi _{\beta}$, and
the $D_{\alpha\beta}$ are, in general, differential operators of
an arbitrary finite order with the coefficients depending on the
variables $\phi_{\alpha}$ and their derivatives (which are also of
a finite order). These operators must satisfy certain conditions
that allow the definition of a Poisson bracket between functionals
of the $\phi _{\alpha}$ (see, e.g., Refs.\ \cite{Olv} and
\cite{GF1}). Here and henceforth a dot denotes differentiation
with respect to the time and there is summation over repeated
indices.

In the case of the source-free electromagnetic field, taking the
components of the electric and the magnetic field as the field
variables $\phi _{\alpha}$, the evolution equations, given by
Maxwell's equations, can be expressed in the form (\ref{eq1}),
without having to introduce the electromagnetic potentials and,
therefore, without having to choose an specific gauge
\cite{Olv}$^{,}$ \cite{GF1}. By contrast, in the standard
Lagrangian formulation for the electromagnetic field, the field
variables are precisely the electromagnetic potentials. In Ref.\
\cite{GF2} a Hamiltonian structure for the linearized Einstein
vacuum field equations is found by using as Hamiltonian density an
analog of the energy of the electromagnetic field. This
Hamiltonian structure involves integral operators. Another
Hamiltonian structure for this linearized theory is found in Ref.\
\cite{RRR} by using another Hamiltonian.

In this paper we show that the evolution equations for the
gravitational field, given by the Einstein vacuum field equations
in an alternative representation derived from that of Ashtekar,
can be expressed in a Hamiltonian form analogous to Eq.\
(\ref{eq1}) with the canonical Hamiltonian structure and in terms
of non canonical variables. This construction is not immediately
obvious. In particular, the covariant derivative in the operators
$D_{\alpha\beta}$, defined below, leads to some difficulties which
will be addressed here. Furthermore, the gauge systems has not
been systematically studied in terms of non canonical variables
(see \cite{MM} and \cite{MM2} for a review).

This paper is organized as follows. We start with a brief summary
of the Hamiltonian formalism for gravity in the ADM variables.
Then we analyze the change of variables leading to the Ashtekar
formalism. In Sect.\ 4 the alternative form of the dynamical
equations of vacuum general relativity is derived. A Poisson
bracket is obtained and it is shown that it yields the expected
relations between the Hamiltonian and any functional of the field.
Then we review the Poisson algebra of the constraints. In Sect.\ 5
we sketch the quantization. We end the paper with the conclusions
and a brief discussion of the prospects related to the alternative
representation.

\section{ADM formalism}
Spacetime can be considered as a 4-manifold $M$, arising as a
result of the time evolution of a three-dimensional space-like
hypersurface $\Sigma$. The manifold $M$ is assumed to be
orientable, and have the global topology $\Sigma\times \Re$. $\Re$
is the real line. We assume that $\Sigma$ is compact without
boundary. The dynamical variables are the Riemannian 3-metric
tensor field $q_{ab}$, and the tensor density field of the
conjugate momenta $p^{ab}$ \cite{ADM}, which are linearly related
to the extrinsic curvature tensor $K_{ab}$ of the hypersurface,
\begin{equation}
p^{ab}=-q^{1/2}(K^{ab}-q^{ab}K), \label{eq2}
\end{equation}
where $q^{ab}$ is the inverse matrix to $q_{ab}$,
$K=q^{ab}K_{ab}$, $q=\det(q_{ab})$, and the latin indices $a, b,
\ldots$ label spatial coordinates and run over the values $1, 2,
3$. These indices are raised and lowered by means of $q_{ab}$.
(See $e.g.$ Ref.\ \cite{Wald} for a nice treatment of this
formulation.)

The dynamic equations are generated by the Hamiltonian
\begin{equation}
    H = \int \left( N \mathcal{H}+ N^{b} \mathcal{H}_{b} \right){\rm d}^{3}x,
\label{eq3}
\end{equation}
which is a linear combination of the (scalar and vectorial)
constraints
\begin{eqnarray}
\mathcal{H} & = &q^{1/2}\left(- ^{3}R + q^{-1}p^{ab}p_{ab} -
\frac{1}{2}q^{-1}p^{2}\right),\label{eq4}     \\
\mathcal{H}^{a} & = &-2q^{1/2}D_{b}\left( q^{-1/2}p^{ab} \right),
\label{eq5}
\end{eqnarray}
and by the canonical Poisson bracket
\begin{equation}
\{q_{ab}(\textbf{x}), p^{cd}(\textbf{y})
\}=\delta^{c}_{(a}\delta^{d}_{b)}\delta^{3}(\textbf{x}-
\textbf{y}), \label{eq6}
\end{equation}
so that
\begin{equation}
\dot{q}_{ab}= \{q_{ab}, H \},\,\,\,\,\ \dot{p}^{ab}= \{p^{ab}, H
\}. \label{eq7}
\end{equation}

In Eqs.\ (\ref{eq4}) and (\ref{eq5}) $p=
p^{a}{}_{a}=p^{ab}q_{ab}$, and $D_{a}$ is the torsion-free
covariant derivative compatible with $q_{ab}$, with Riemann
curvature tensor $2D_{[a}D_{b]}v_{c}\equiv
{}^{3}R_{abc}{}^{d}v_{d}$, where $v_{a}$ is an arbitrary covector
on $\Sigma$. $^{3}R$ is the Ricci scalar of this curvature. The
scalar $N$ is known as the lapse and $N^{a}$ is a vector on
$\Sigma$ and is usually referred to as the shift vector; they
should be viewed as Lagrange multipliers.

Explicitly the dynamical equations (\ref{eq7}) are given by
\begin{equation}
\dot{q}_{ab}= \frac{\delta H}{\delta
p^{ab}}=2q^{-1/2}N\left(p_{ab}-\frac{1}{2}q_{ab}p\right)+2D_{(a}N_{b)},\label{eq8}
\end{equation}and
\begin{eqnarray}
\dot{p}^{ab}= -\frac{\delta H}{\delta q_{ab}} & = &
-Nq^{1/2}\left(^{3}R^{ab}-\frac{1}{2}{}^{3}Rq^{ab}\right){} \nonumber\\
& &+q^{1/2}D_{c}\left(q^{-1/2}N^{c}p^{ab}\right) {} \nonumber\\
& &
-2Nq^{-1/2}\left(p^{ac}p_{c}{}^{b}-\frac{1}{2}pp^{ab}\right){} \nonumber\\
& &-2p^{c(a}D_{c}N^{b)}{} \nonumber\\& &+
\frac{1}{2}Nq^{-1/2}q^{ab}\left(p_{cd}p^{cd}-\frac{1}{2}p^{2}\right)
{} \nonumber\\& & + q^{1/2}(D^{a}D^{b}N-q^{ab}D^{c}D_{c}N) ,
\label{eq9}
\end{eqnarray}
where boundary terms have been ignored. Equations (\ref{eq4}),
(\ref{eq5}), (\ref{eq8}) and (\ref{eq9}) are equivalent to the
vacuum Einstein equation, $R_{\alpha\beta}=0$
($\alpha,\beta=0,1,2,3$, here). One can explicitly reconstruct the
four-dimensional space-time geometry in arbitrary coordinates
$X^{\alpha}$. For a more thorough treatment of this Hamiltonian
formalism, see Ref.\ \cite{Kuch}.

\section{Ashtekar formalism}
The original literature on Ashtekar's variables uses the language
of $SU(2)$ spinors. We have preferred to avoid this language, and
use $SO(3)$-valued variables. The translation from $SO(3)$-valued
variables to $SU(2)$ spinors is illustrated clearly in Ref.\
\cite{Ash3}. Moreover, in this section our convention is closer to
that of Ref.\ \cite{Cap}.

Instead of the metric tensor $q_{ab}$ we introduce the triad
$e_{a}{}^{i}$, such that the spatial metric is given by
\begin{equation}
q_{ab}=e_{a}{}^{i}e_{b}{}^{j}\delta_{ij}=e_{a}{}^{i}e_{bi}
\label{eq10}
\end{equation}
Latin indices $i,j,...=1,2,3$, from the middle of the alphabet,
are $SO(3)$ indices. They are raised and lowered with the
Kronecker delta $\delta^{ij}$. The inverse matrices to the triad
are denoted by $e^{a}{}_{i}$, hence,
$e_{a}{}^{i}e^{b}{}_{i}=\delta_{a}^{b}$, and
$e_{a}{}^{i}e^{a}{}_{j}=\delta_{i}^{j}$. Since
$q^{cb}q_{ba}=q^{cb}e_{b}{}^{i}e_{a}{}^{i}=\delta_{a}^{c}$, the
inverse matrix can be obtained by raising the index with the help
of $q^{cb}$, $e^{c}{}_{i}=e^{ci}=q^{cb}e_{b}{}^{i}$. The position
of the internal index $i$ is irrelevant. It is also not difficult
to verify that
\begin{equation}
q^{ab}=e^{ai}e^{bi},\,\,\,\,\,
q=\det(e_{a}{}^{i}e_{bi})=(\det(e_{a}{}^{i}))^{2}\equiv e^{2}.
\label{eq11}
\end{equation}

Let us introduce the momenta $p^{a}{}_{i}$ conjugate to the triad.
They satisfy the equations
\begin{equation}
\left\{e_{a}{}^{i}(\textbf{x}), p^{b}{}_{j}(\textbf{y})
\right\}=\delta^{b}_{a}\delta^{i}_{j}\delta^{3}(\textbf{x}-
\textbf{y}), \label{eq12}
\end{equation}
and can be easily related to the momenta $p^{ab}$ by means of
\begin{equation}
p^{a}{}_{i} = 2p^{ab}e_{bi}. \label{eq13}
\end{equation}

It now turns out that part of the Poisson brackets for the ADM
variables has been modified:
\begin{equation}
\left\{q_{ab}(\textbf{x}), p^{cd}(\textbf{y})
\right\}=\delta^{c}_{(a}\delta^{d}_{b)}\delta^{3}(\textbf{x}-
\textbf{y}),\,\,\,\, \left\{q_{ab}(\textbf{x}), q_{cd}(\textbf{y})
\right\}=0, \label{eq14}
\end{equation}
while
\begin{eqnarray}
\left\{p^{ab}(\textbf{x}), p^{cd}(\textbf{y})
\right\}&=&\frac{1}{4}(q^{ac}J^{bd}+q^{ad}J^{bc}+
q^{bc}J^{ad}{}  \nonumber\\
& & +q^{bd}J^{ac})\delta^{3}(\textbf{x}- \textbf{y}),\label{eq15}
\end{eqnarray}
where
\begin{equation}
J^{ab}=\frac{1}{4}(e^{ai}p^{b}{}_{i}-e^{bi}p^{a}{}_{i})= J^{[ab]}.
\label{eq16}
\end{equation}
To preserve the correspondence between Poisson structures, one has
to impose the three constraints $J^{ab}=0$, which also ensures the
conservation of the number of degrees of freedom (a symmetric
tensor $q_{ab}$ is defined by six numbers at each point, while the
triad matrix $e_{a}{}^{i}$ contains nine independent components).
These additional constraints generate $SO(3)$ rotations (which
leave $q_{ab}$ invariant) and can be represented equivalently in
the form (see $e.g.$ Ref.\ \cite{Cap})
\begin{equation}
\mathcal{J}^{i} = \epsilon^{ijk}p^{a}{}_{j}e_{ak}=0,\label{eq17}
\end{equation}
where $\epsilon^{ijk}$ is the totally antisymmetric Levi-Civita
symbol ($\epsilon^{123}= 1$).

Thus, the constraint $\mathcal{J}^{i}$ implements the condition
that $p_{ab}$, considered now as a derived quantity, is symmetric
\begin{equation}
p^{ab}=\frac{1}{4}(p^{a}{}_{i}e^{b}{}_{i}+p^{b}{}_{i}e^{a}{}_{i}).
\label{eq18}
\end{equation}

In terms of $(e_{a}{}^{i}, p^{a}{}_{i})$, the Hamiltonian becomes
\begin{equation}
H=\int\left(N\mathcal{H}+N^{a}\mathcal{H}_{a}+N_{i}\mathcal{J}^{i}\right){\rm
d}^{3}x, \label{eq19}
\end{equation}
where $\mathcal{H}$, $\mathcal{H}_{a}$ are given by (\ref{eq4}),
(\ref{eq5}), with $q_{ab}$ and $p^{ab}$ considered here as derived
quantities, and we have annexed the additional constraint with the
Lagrange multiplier $N_{i}$.

Clearly, the choice of $(e_{a}{}^{i}, p^{a}{}_{i})$ as the
canonical variables is not unique. In view of the transition to
the Ashtekar variables that we make below, it is more convenient
to use the variables $(E^{ai}, K_{a}{}^{i})$ defined by
\begin{equation}
E^{a}{}_{i}\equiv ee^{a}{}_{i},\,\,\,\,\, K_{a}{}^{i}\equiv
K_{ab}e^{b}{}_{i}+ J_{ab}e^{b}{}_{i},\label{eq20}
\end{equation}
where $K_{ab}=K_{(ab)}$ is the extrinsic curvature, and $J_{ab}$
is given by (\ref{eq16}). Then
\begin{equation}
\{E^{a}{}_{i}(\textbf{x}),
K_{b}{}^{j}(\textbf{y})\}=\frac{1}{2}\delta^{a}_{b}\delta^{j}_{i}
\delta^{3}(\textbf{x}-\textbf{y}), \label{eq21}
\end{equation}
\begin{equation}
\{E^{a}{}_{i}(\textbf{x}),E^{b}{}_{j}(\textbf{y})\}= 0,\,\,\,\,\,
\{K_{a}{}^{i}(\textbf{x}),K_{b}{}^{j}(\textbf{y})\}= 0.
\label{eq22}
\end{equation}

In \cite{Ash1} Ashtekar proposed a transformation that allowed one
to represent the density of the gravitational Hamiltonian as a
polynomial in canonical variables. The transformation is
canonical, up to a surface term. Since we are considering a closed
$\Sigma$ without boundary, this term vanishes.

Ashtekar also introduced a complex parametrization in which the
new variables are represented as
\begin{equation}
A_{a}{}^{i}= \frac{1}{2}\epsilon^{ijk}e^{b}{}_{k}D_{a}e_{b}{}^{j}
+ i K_{a}{}^{i}.\label{eq23}
\end{equation}
In this parametrization, we have
\begin{equation}
\{E^{a}{}_{i}(\textbf{x}),
A_{b}{}^{j}(\textbf{y})\}=i\delta^{a}_{b}\delta^{j}_{i}
\delta^{3}(\textbf{x}-\textbf{y}), \label{eq24}
\end{equation}
\begin{equation}
\{E^{a}{}_{i}(\textbf{x}),E^{b}{}_{j}(\textbf{y})\}= 0,\,\,\,\,\,
\{A_{a}{}^{i}(\textbf{x}),A_{b}{}^{j}(\textbf{y})\}= 0.
\label{eq25}
\end{equation}
For any two functionals in phase space $F(E, A)$, $G(E, A)$, the
Poisson bracket is thus given by
\begin{equation}
\{ F, G \} \equiv i \int \left( \frac{\delta F}{\delta
E^{a}{}_{i}}\frac{\delta G}{\delta A_{a}{}^{i}} - \frac{\delta
F}{\delta A_{a}{}^{i}}\frac{\delta G}{\delta
E^{a}{}_{i}}\right){\rm d}^{3}x.\label{eq25a}
\end{equation}

Changing the variables in the Hamiltonian leads to the expression
\begin{equation}
H = i\int\left(-\frac{i}{2}\mathcal{N} \mathcal{S}+
\frac{1}{2}N^{a}\mathcal{V}_{a}+ N^{i}\mathcal{G}_{i}\right){\rm
d}^{3}x, \label{eq26}
\end{equation}
where
\begin{eqnarray}
\mathcal{G}_{i}(A, E) & \equiv & \mathcal{D}_{a}E^{a}{}_{i} \equiv i\epsilon^{abc}J_{ab}e_{c}{}^{i}=0 , \label{eq27}     \\
\mathcal{V}_{a}(A, E) & \equiv & E^{b}{}_{i}F_{ab}{}^{i} = 0, \label{eq28}       \\
\mathcal{S}(A, E) & \equiv &
E^{a}{}_{i}E^{b}{}_{j}F_{abk}\epsilon^{ijk} = 0. \label{eq29}
\end{eqnarray}
are the (Gauss, vectorial and scalar) constraints,
$\mathcal{N}=e^{-1}N$ and $\epsilon^{abc}$ is the totally
antisymmetric Levi-Civita symbol ($\epsilon^{123}= 1$) .

The covariant derivative $\mathcal{D}_{a}$ is defined by
\begin{equation}
\mathcal{D}_{a}v_{i}\equiv \partial_{a}v_{i} +
\frac{1}{2}\epsilon_{ijk}A_{a}{}^{j}v^{k}. \label{eq30}
\end{equation}

The curvature of the connection $A_{a}{}^{i}$ can be found from
\begin{equation}
2\mathcal{D}_{[a}\mathcal{D}_{b]}v_{i}= \frac{1}{2}
\epsilon_{ijk}F_{ab}{}^{j}v^{k}, \label{eq31}
\end{equation}
hence
\begin{equation}
F_{ab}{}^{i} = \partial_{a}A_{b}{}^{i} -
\partial_{b}A_{a}{}^{i} +
\frac{1}{2}\epsilon^{i}{}_{jk}A_{a}{}^{j}A_{b}{}^{k}. \label{eq32}
\end{equation}

The evolution equations for the canonical variables are obtained
taking the Poisson bracket of the variables with the Hamiltonian
(\ref{eq26}), and, neglecting boundary terms, they are given by
\begin{eqnarray}
\dot{A}_{a}{}^{i}(x) & = & \{A_{a}{}^{i}, H\} =
-i\mathcal{N}\epsilon_{ijk}E^{bj}F_{ab}{}^{k} {}\nonumber\\&&+
\frac{1}{2}N^{b}F_{ba}{}^{i} , \label{eq33}     \\
\dot{E}^{a}{}_{i}(x) & = & \{E^{a}{}_{i}, H\} =
i\mathcal{D}_{b}(\mathcal{N}\epsilon_{ijk}E^{[a \mid j
\mid}E^{b]k}){}\nonumber\\&& +
\mathcal{D}_{b}(N^{[b}E^{a]}{}_{i}). \label{eq34}
\end{eqnarray}
A simplification is evident in the equations of motion.

\section{The alternative Hamiltonian formulation}

In this section, we shall review the alternative Hamiltonian
formulation for general relativity that emerges from Ashtekar's
canonical gravity and from the Hamiltonian formulation outlined in
the Introduction which is wider than the one derived from the
Lagrangian formulation. This Hamiltonian formulation is based in
the fact that the time evolution of the field variables
$\phi_{\alpha}$ can be written in the form (\ref{eq1}). Clearly,
for a candidate Hamiltonian operator $D_{\alpha\beta}$ [$cf.$ Eq.\
(\ref{eq1})], the correct expression for the corresponding Poisson
bracket has the form
\begin{equation}
\left\{ F, G \right\} = \int  \frac{\delta F}{\delta
\phi_{\alpha}}D_{\alpha\beta}\frac{\delta G}{\delta \phi_{\beta}}
{\rm d}^{3}x,\label{eq35}
\end{equation}
where $F$ and $G$ are functionals. Of course, the Hamiltonian
operator $D_{\alpha\beta}$ must satisfy certain further
restrictions in order for (\ref{eq35}) to be a true Poisson
bracket. The condition that
$D_{\alpha\beta}=-\overline{D^{\dag}_{\beta\alpha}}$, where
$D^{\dag}_{\alpha\beta}$ is the adjoint of $D_{\alpha\beta}$ and
the bar denotes complex conjugation, is equivalent to the
antisymmetry of the Poisson bracket ($i.e.$, $ \{F, G \} = - \{G,
F \}$). The other condition on the Poisson bracket is the Jacobi
identity; when the $D_{\alpha\beta}$ are constants, this condition
is automatically satisfied, but in other cases one has to verify
that this identity is satisfied \cite{Olv}.

The Poisson bracket (\ref{eq35}), while is formally correct, fails
to incorporate boundary effects, and needs to be slightly modified
when discussing solutions over bounded domains (see $e.g.$ Refs.
\cite{LMMR} and \cite{Sol} for this point). However, we have
assumed that $\Sigma$ is compact without boundary here.

Using the fact that
\begin{equation}
\phi_{\alpha}(\textbf{y}, t) = \int \delta_{\alpha\beta}
\delta^{3}(\textbf{x}-\textbf{y})\phi_{\beta}(\textbf{x},
t)d^{3}x\label{eq36}
\end{equation}
it follows that
\begin{equation}
\frac{ \delta \phi_{\alpha}(\textbf{y}, t)}{ \delta
\phi_{\beta}(\textbf{x}, t)} = \delta_{\alpha\beta}
\delta^{3}(\textbf{x}-\textbf{y}), \label{eq37}
\end{equation}
therefore, from Eq.\ (\ref{eq35}), one gets
\begin{equation}
\{\phi_{\alpha}(\textbf{x}, t), \phi_{\beta}(\textbf{y}, t)\} =
D_{\alpha\beta}\delta^{3}(\textbf{x}-\textbf{y}). \label{eq38}
\end{equation}
In the simplest case of the canonical variables
$\phi_{\alpha}=(q^{i}, p_{i})$, like the ADM and Ashtekar
variables, the operator $D \equiv (D_{\alpha\beta})$ is the
antisymmetric matrix
\begin{equation}
D = \left(%
\begin{array}{cc}
  0 & \delta^{i}_{j} \\
  - \delta^{i}_{j} &  0 \\
\end{array}%
\right). \label{eq39}
\end{equation}
Of course, the ADM and Ashtekar variables have two indices and the
fundamental canonical Poisson brackets are given by (\ref{eq6})
and (\ref{eq24}).

Coming back, now, to the case of general relativity, we wish to
use $E^{a}{}_{i}$ and the ``magnetic field" $B^{a}{}_{i}$ as new
variables, rather than $E^{a}{}_{i}$ and $A_{a}{}^{i}$. First, we
define
\begin{equation}
B^{a}{}_{i}\equiv \epsilon^{abc}F_{bci} =
\epsilon^{abc}\left(2\partial_{b}A_{ci}+\frac{1}{2}\epsilon^{ijk}A_{bj}A_{ck}\right),\label{eq40}
\end{equation}
then one finds that
\begin{equation}
\dot{B}^{a}{}_{i} = \{B^{a}{}_{i}, H\}=2
\epsilon^{abc}\mathcal{D}_{b}\dot{A}_{ci}.\label{eq41}
\end{equation}

From Eq.\ (\ref{eq40}) it should be stated that given $A_{ai}$,
one can calculate $B^{a}{}_{i}$, but, can this relation be
inverted? The answer is no in general.

Let us consider the possibility of describing the configuration
space of the system using $B^{a}{}_{i}$ rather than $A_{ai}$,
which will be necessary in order to write the connection from the
covariant derivative in terms of $B^{a}{}_{i}$ and their partial
derivatives.

For the non-Abelian theory is generically possible to view the
Bianchi identity
\begin{equation}
\mathcal{D}_{a}B^{ai}=\partial_{a}B^{ai}+\frac{1}{2}\epsilon^{ijk}A_{aj}B^{a}{}_{k}=0,
\end{equation}
as a linear relation between $B^{ai}$ and $A_{ai}$ [which is
compatible with (\ref{eq40})] to be solved for $A_{ai}$, thus,
$B^{ai}$ can be used as a variable.

Now, it is possible that two or more gauge inequivalent
non-Abelian potentials $A_{ai}$ generate the same field $B^{ai}$,
which is known as the Wu-Yang ambiguity \cite{W-Y}. But there
exist some examples in the context of $SU(2)$ gauge theories
\cite{Freed}, in which the correspondence between $A_{ai}$ and
$B^{ai}$ modulo gauge is made, but some conditions on $B^{ai}$ are
necessary (see also \cite{W-Y}). (One condition is that the
$3\times3$ matrix $B^{ai}$ satisfies $\det B^{ai} \neq 0$.)

Therefore, in that follows we will suppose that it is possible to
write $A=A(B)$, (we can consider the conditions on $B^{ai}$ given
in \cite{Freed} for $SO(3)$ theories, for instance).

On the other hand, we can express the evolution equations in terms
of the variables $E^{a}{}_{i}$ and $B^{a}{}_{i}$ only. Equation
(\ref{eq34}) remains unchanged and Eq.\ (\ref{eq33}) can be
rewritten as
\begin{equation}
\dot{A}_{ai}(E, B)  =
-\frac{i}{2}\mathcal{N}\epsilon_{ijk}E^{bj}\epsilon_{abc}B^{ck} +
\frac{1}{4}N^{b}\epsilon_{bac}B^{c}{}_{i} , \label{eq42}
\end{equation}
where we have used the fact that $2F_{ab}{}^{i}=
\epsilon_{abc}B^{ci}$ [$cf$.\ Eq.\ (\ref{eq40})]. Thus, we have
that [$cf$.\ Eq.\ (\ref{eq41})]
\begin{equation}
2 \epsilon^{abc}\mathcal{D}_{b}\dot{A}_{ci}=
-2i\epsilon_{ijk}\mathcal{D}_{b}(\mathcal{N}E^{[a\mid j
\mid}B^{b]k}) - \mathcal{D}_{b}(N^{[a}B^{b]}{}_{i}).\label{eq43}
\end{equation}
Therefore, we have an alternative set of equations of evolution
for the gravitational field equivalent to Eqs.\ (\ref{eq34}) and
(\ref{eq33}), given by
\begin{eqnarray}
\dot{E}^{a}{}_{i} & = &
i\mathcal{D}_{b}(\mathcal{N}\epsilon_{ijk}E^{[a \mid j
\mid}E^{b]k}) + \mathcal{D}_{b}(N^{[b}E^{a]}{}_{i}),  \label{eq44}
\\
\dot{B}^{a}{}_{i} & = &
-2i\mathcal{D}_{b}(\mathcal{N}\epsilon_{ijk}E^{[a\mid j
\mid}B^{b]k}) - \mathcal{D}_{b}(N^{[a}B^{b]}{}_{i}). \label{eq45}
\end{eqnarray}
These equations are more symmetric than Eqs.\ (\ref{eq34}) and
(\ref{eq33}), and in some sense they are analogous to the Maxwell
equations. Therefore, in terms of the variables $E^{a}{}_{i}$ and
$B^{a}{}_{i}$, which are not canonical, the equations of evolution
for vacuum general relativity take an interesting form. However,
this is not sufficient. What is needed is a Hamiltonian structure
that defines a Poisson bracket and a Hamiltonian which involves
the constraints and generates the evolution equations (\ref{eq44})
and (\ref{eq45}). We also need that the constraint algebra of the
constraints closes.

\subsection{Hamiltonian and Hamiltonian structure}

In order to express the alternative set of evolution equations in
the Hamiltonian form (\ref{eq1}), we introduce the Hamiltonian
\begin{equation}
H = i\int {\rm d}^{3}x \left(-\frac{i}{2}\mathcal{N}\mathcal{S}+
\frac{1}{2}N^{a}\mathcal{V}_{a}+ N^{i}\mathcal{G}_{i}\right)
\label{eq46}
\end{equation}
where, now,
\begin{eqnarray}
\mathcal{V}_{a}(E, B) & \equiv & \frac{1}{2}\epsilon_{abc}E^{b}{}_{i}B^{ci} = 0, \label{eq47}       \\
\mathcal{S}(E, B) & \equiv &
\frac{1}{2}\epsilon_{abc}E^{a}{}_{i}E^{b}{}_{j}B^{c}{}_{k}\epsilon^{ijk}=
0, \label{eq47a}       \\
\mathcal{G}_{i}(E, B)& \equiv & \mathcal{D}_{a}E^{a}{}_{i}
\label{eq48},
\end{eqnarray}
are the constraints. The Hamiltonian (\ref{eq46}) is the same of
Ashtekar [$cf.$ Eq.\ (\ref{eq26})], but, now, in the Gauss
constraint we consider $A=A(B)$.

On the other hand, Eqs.\ (\ref{eq44}) and (\ref{eq45}) can be
written in the Hamiltonian form
\begin{equation}
\dot{E}^{a}{}_{i}= \mathcal{D}^{ab}{}_{ij} \frac{\delta H}{\delta
B^{b}{}_{j}},\,\,\,\,\,\,\, \dot{B}^{a}{}_{i}=
-\mathcal{D}^{ab}{}_{ij} \frac{\delta H}{\delta E^{b}{}_{j}}
\label{eq51}
\end{equation}
where
\begin{equation}
\mathcal{D}^{ab}{}_{ij} \equiv-2i
\epsilon^{abc}\mathcal{D}_{c}\delta_{ij}\equiv
-2i\epsilon^{abc}\left(
\partial_{c}\delta_{ij}+\frac{1}{2}\epsilon_{ikl}A_{c}{}^{k}\delta^{l}_{j}\right) \label{eq52}
\end{equation}
and $H$ is given by (\ref{eq46}).

In this case, the matrix differential operator $D=
(D_{\alpha\beta})$ [$cf.$ Eq.\ (\ref{eq1})] can be seen in a
schematic form (forgetting for a moment the internal indices $i,
j$) as
\begin{equation}
D=(D_{\alpha\beta}) = \left(%
\begin{array}{cc}
  0 & \epsilon^{abc}\mathcal{D}_{c} \\
  - \epsilon^{abc}\mathcal{D}_{c} &  0 \\
\end{array}%
\right), \label{eq53}
\end{equation}
($\alpha,\beta$ = 1, 2, \ldots, 6 here).

Making use of the $\mathcal{D}^{ab}{}_{ij}$ given by Eq.\
(\ref{eq52}), a Poisson bracket between any pair of functionals of
the field $F(E, B)$ and $G(E, B)$ can be defined as
\begin{equation}
\{ F, G \}_{n}\equiv \int \left( \frac{\delta F}{\delta
E^{a}{}_{i}}\mathcal{D}^{ab}{}_{ij}\frac{\delta G}{\delta
B^{b}{}_{j}} - \frac{\delta F}{\delta
B^{a}{}_{i}}\mathcal{D}^{ab}{}_{ij}\frac{\delta G}{\delta
E^{b}{}_{j}}\right){\rm d}^{3}x,\label{eq54}
\end{equation}
where the subscript $n$ (\emph{non} canonical variables) is
introduced to distinguish it from the canonical Poisson bracket.

Integrating by parts one can see that the bracket (\ref{eq54}) is
antisymmetric up to a surface term; since we are considering a
closed $\Sigma$ without boundary this term vanishes. Equivalently,
the matrix differential operator $D$ is skew-adjoint \cite{Olv}
($i.e.$ $D^{\dag}=-D$).

In order to prove the Jacobi identity for this Poisson bracket we
will use the methods of functional multi-vectors given in Ref.\
\cite{Olv}. In such case the Jacobi identity is equivalent to the
condition that the functional tri-vector
\begin{equation}
\Psi \equiv \frac{1}{2} \int [\theta \wedge \textrm{pr}
\textbf{v}_{D \theta}(D)\wedge \theta]{\rm d}v\label{eq55}
\end{equation}
vanishes. Here $\theta$ is some functional uni-vector and
$\textrm{pr}\textbf{v}_{D \theta}(D)$ is the prolongation of the
evolutionary vector field $\textbf{v}_{D\theta}$ (with
characteristic $D \theta$) acting on the skew-adjoint differential
operator $D$. For the computation of
$\textrm{pr}\textbf{v}_{D\theta}(D)$ it is necessary to write $D$
in terms of the field variables, $E^{a}{}_{i}$ and $B^{a}{}_{i}$,
only, $i.e.$ to write $A_{a}{}^{i}$ in terms of $B^{a}{}_{i}$. In
any (possible) case, the relation $A=A(B)$ does not involve
differential operators, then $\textrm{pr}\textbf{v}_{D\theta}(D)$
turns out to be some uni-vector, $\vartheta$, that does not
involve differential operators. Thus
\begin{equation}
\Psi = \frac{1}{2} \int [\theta\wedge \vartheta\wedge\theta]{\rm
d}v = 0\label{eq59}
\end{equation}
(by the antisymmetry of the wedge product). Hence, the operators
$\mathcal{D}^{ab}{}_{ij}$ define a Hamiltonian structure, or,
equivalently, a Poisson bracket.

Also one can see that the antisymmetry and the Jacobi identity of
the Poisson bracket (\ref{eq54}) follows from the fact that the
Hamiltonian structure is the canonical one, $i.e$, from the
canonical Poisson bracket (\ref{eq25a}), by using the fact of that
\begin{equation}\label{CHR}
    \frac{\delta}{\delta
    A_{a}{}^{i}}=2\epsilon^{abc}\mathcal{D}_{b}\frac{\delta}{\delta
    B^{c}{}_{i}},
\end{equation}
which follows from the chain rule, one can see that (integrating
by parts)
\begin{equation}\label{}
\{F, G\}=\{F, G\}_{n}
\end{equation}
[$cf.$ Eqs.\ (\ref{eq25a}) and (\ref{eq54})]. Therefore, the
Hamiltonian structure is the canonical one, only the variables are
new. Thus, in that follows, we will use the subscript $n$ in order
to point out that we are using non canonical variables.

The new variables satisfy the Poisson brackets relations
\begin{equation}
\{E^{a}{}_{i}(\textbf{x}),E^{b}{}_{j}(\textbf{y}) \}_{n}=0,
\,\,\,\, \{B^{a}{}_{i}(\textbf{x}),B^{b}{}_{j}(\textbf{y})
\}_{n}=0 \label{eq603}
\end{equation}
and
\begin{eqnarray}
&& \{E^{a}{}_{i}(\textbf{x}), B^{b}{}_{j}(\textbf{y}) \}_{n}  =
-2i\epsilon^{abc}\mathcal{D}_{c} \delta_{ij}
\delta^{3}(\textbf{x}-\textbf{y}){}\nonumber\\
&&=-2i\epsilon^{abc}
\left(\partial_{c}\delta_{ij}+\frac{1}{2}\epsilon_{ilj}A_{c}{}^{l}
\right)\delta^{3}(\textbf{x}-\textbf{y}). \label{eq60}
\end{eqnarray}

The Poisson bracket (\ref{eq54}) yields the expected relations
between the Hamiltonian and any functional of the field. If $F(E,
B)$ is any functional of the field that does not depend explicitly
on the time then Eqs.\ (\ref{eq54}) and (\ref{eq51}) give
\begin{eqnarray}
\{ F, H \}_{n} & = & \int \left( \frac{\delta F}{\delta
E^{a}{}_{i}}\mathcal{D}^{ab}{}_{ij}\frac{\delta H}{\delta
B^{b}{}_{j}} - \frac{\delta F}{\delta
B^{a}{}_{i}}\mathcal{D}^{ab}{}_{ij}\frac{\delta H}{\delta
E^{b}{}_{j}}\right){\rm d}^{3}x {}\nonumber\\
& = & \int \left( \frac{\delta F}{\delta E^{a}{}_{i}}
\dot{E}^{a}{}_{i} + \frac{\delta F}{\delta B^{a}{}_{i}}
\dot{B}^{a}{}_{i}\right){\rm d}^{3}x = \dot{F}, \label{eq62}
\end{eqnarray}
$i.e.$, $H$ generates time translations.

\subsection{Constraint Poisson algebra}
The geometrical interpretation of the constraints is made easier
by integrating them over the enteire spatial slice as
\begin{equation}
C_{N^{i}}\equiv \int {\rm
d}^{3}xN^{i}\mathcal{G}_{i}(E,B),\label{eq61}
\end{equation}
\begin{equation}
    C'_{N^{a}}\equiv \frac{1}{2}\int {\rm d}^{3}xN^{a}\mathcal{V}_{a}(E,B),\label{eq61b}
\end{equation}
\begin{equation}
    C_{\mathcal{N}}\equiv -\frac{i}{2}\int {\rm d}^{3}x\mathcal{N}\mathcal{S}(E,B).\label{eq61c}
\end{equation}

First we consider the Gauss constraint. By using Eq.\ (\ref{CHR})
one has that
\begin{eqnarray}
\{E^{a}{}_{i}, C_{N^{i}}
\}_{n}&=&\frac{i}{2}\epsilon^{ijk}E^{a}{}_{j}N^{k}, {}\nonumber\\
\{B^{a}{}_{i}, C_{N^{i}}
\}_{n}&=&\frac{i}{2}\epsilon^{ijk}B^{a}{}_{j}N^{k}, \label{}
\end{eqnarray}
$i.e.$, Gauss constraint generates internal rotations on the
variables.

Next, let us consider the vector constraint (\ref{eq61b}). To
obtain the generator of spatial diffeomorphisms, one has to add to
Eq.\ (\ref{eq61b}) a multiple of Eq.\ (\ref{eq61}). Given a shift
vector $N^{a}$, let us set
\begin{equation}
    C_{N^{a}}\equiv \frac{1}{2}\int {\rm d}^{3}xN^{a}\mathcal{V}_{a}(E,B)
    -\frac{1}{2}\int {\rm d}^{3}xN^{b}A_{b}{}^{i}\mathcal{D}_{a}E^{a}{}_{i}. \label{eq61d}
\end{equation}
Then, by using the Poisson bracket (\ref{eq54}) and Eq.\
(\ref{CHR}), one can show that
\begin{eqnarray}
\{E^{a}{}_{i}, C_{N^{a}}
\}_{n}&=&-\frac{i}{2}\pounds_{N^{a}}E^{a}{}_{i},
{}\nonumber\\\{B^{a}{}_{i}, C_{N^{a}}
\}_{n}&=&-\frac{i}{2}\pounds_{N^{a}}B^{a}{}_{i} \label{eq}
\end{eqnarray}
where $\pounds_{N^{a}}$ is the Lie derivative, with respect to the
vector field $N^{a}$. Thus, $C_{N^{a}}$ does indeed generate
diffeomorphisms along the vector field $N^{a}$.

Finally, let us consider the scalar constraint (\ref{eq61c}). In
this case $C_{\mathcal{N}}$ has the following Poisson bracket
relations
\begin{eqnarray}
\{E^{a}{}_{i}, C_{\mathcal{N}}
\}_{n}&=&\mathcal{D}_{b}(\mathcal{N}\epsilon_{ijk}E^{[a|j|}E^{b]k}),{}\nonumber\\
 \{B^{a}{}_{i}, C_{\mathcal{N}}
\}_{n}&=&-2\mathcal{D}_{b}(\mathcal{N}\epsilon_{ijk}E^{[a|j|}B^{b]k})\label{e}
\end{eqnarray}
[$cf.$ Eqs.\ (\ref{eq44}) and (\ref{eq45})].

We can now compute the Poisson bracket algebra. One readily
obtains
\begin{eqnarray}\label{}
\{C_{N^{i}} , C_{M^{i}}
\}_{n}&=&-\frac{i}{2}C_{\epsilon_{ijk}N^{j}M^{k}} \label{}{}\nonumber\\
\{C_{N^{a}} , C_{M^{a}}
\}_{n}&=&\frac{i}{2}C_{\pounds_{N^{a}}M^{a}} \label{}{}\nonumber\\
\{C_{N^{a}} , C_{N^{i}}
\}_{n}&=&\frac{i}{2}C_{\pounds_{N^{a}}N^{i}} \label{}{}\nonumber\\
\{C_{N^{a}} , C_{\mathcal{N}}
\}_{n}&=&\frac{i}{2}C_{\pounds_{N^{a}}\mathcal{N}} \label{}{}\nonumber\\
\{C_{N^{i}} , C_{\mathcal{N}} \}_{n}&=&0 \label{}{}\nonumber\\
\{C_{\mathcal{N}} , C_{\mathcal{M}}
\}_{n}&=&-iC_{K^{a}}-\frac{i}{2}C_{K^{a}A_{a}{}^{i}},
\end{eqnarray}
 where the vector $K$ is defined by
$K^{a}=E^{a}{}_{i}E^{bi}(\mathcal{N}\partial_{b}\mathcal{M}
-\mathcal{M}\partial_{b}\mathcal{N})$. Here we clearly see that
the algebra closes. In the Dirac terminology, the algebra is first
class.

The counting of the degrees of freedom is done as in the
Ashtekar's case. We have a 18-dimensional phase space. In that
space we have seven constraints and we can fix seven gauge
conditions. We are therefore left with a four-dimensional
constraint-free space, which gives two degrees of freedom. Note
that the Bianchi identity, $\mathcal{D}_{a}B^{a}{}_{i}=0$, does
not reduce the dimension of the phase space, since it is not a
constraint.

\section{Sketching quantization}

Now we sketch briefly the quantization by using the new
Hamiltonian approach (we will not go into the details). In
principle, it is entirely straightforward to quantize the theory.
However, quantum gravity is still poorly understood, and we will
be sketching a program that people hoped would lead to a theory of
quantum gravity, but which has technical complications.

Quantization requires us to replace the variables ($E$ and $B$ in
this case) by operators that act on the space of states of the
theory. The wave-functionals that are annihilated by the
constraints are the physical states of the theory. Notice that we
do not yet have a Hilbert space. One needs to introduce an inner
product on the space of physical states in order to compute
expectation values and make physical predictions. We will ignore
these questions for now and think of the states as arbitrary
functionals $\psi[B]$.

We replace the classical variables $E^{a}{}_{i}$ and $B^{a}{}_{i}$
by the operators
\begin{equation}
\hat{E}^{a}{}_{i}(\textbf{x})\psi[B]=
2i\epsilon^{abc}\mathcal{D}_{b}\frac{\delta}{\delta
B^{c}{}_{i}(\textbf{x})}\psi[B], \label{eq605}
\end{equation}
and
\begin{equation}
\hat{B}^{a}{}_{i}(\textbf{x})\psi[B]=
B^{a}{}_{i}(\textbf{x})\psi[B], \label{eq605}
\end{equation}
which have commutation relations
\begin{equation}
[\hat{E}^{a}{}_{i}(\textbf{x}),\hat{E}^{b}{}_{j}(\textbf{y}) ]=0,
\,\,\,\,
[\hat{B}^{a}{}_{i}(\textbf{x}),\hat{B}^{b}{}_{j}(\textbf{y}) ]=0,
\label{eq606}
\end{equation}
\begin{eqnarray}
[\hat{E}^{a}{}_{i}(\textbf{x}),\hat{B}^{b}{}_{j}(\textbf{y}) ] & =
& 2i\epsilon^{abc}\mathcal{D}_{c} \delta_{ij}
\delta^{3}(\textbf{x}-\textbf{y}), \label{eq607}
\end{eqnarray}
analogous to the classical Poisson brackets relations
(\ref{eq603}) and (\ref{eq60}).

With these operators at hand one can promote the constraints
formally to operator equations if one picks a factor ordering.
There are two factor orderings which have been explored: with the
$E's$ either to the right \cite{Smo} or the left of the $B's$.

The problem, now, is to find the physical state space of functions
$\psi[B]$ that satisfy the constraints in quantum form.

For the factor ordering with $E's$ to the left there exist the
Chern-Simons state $\psi_{CS}[A]$ (in Ashtekar representation),
which satisfies
\begin{equation}
\hat{\mathcal{G}}_{i}\psi_{CS}=\hat{\mathcal{V}}_{a}\psi_{CS}=\hat{\mathcal{S}}\psi_{CS}=0,
\label{eq614}
\end{equation}
whenever the cosmological constant $\Lambda$ is nonzero. In this
case the Gauss and vectorial constraints are unchanged, but the
scalar constraint becomes \cite{Gamb}
\begin{eqnarray}
\hat{\mathcal{S}}_{\Lambda}(E, B) & = &
\frac{1}{2}\epsilon_{abc}\hat{E}^{a}{}_{i}\hat{E}^{b}{}_{j}\hat{B}^{c}{}_{k}\epsilon^{ijk} {}  \nonumber\\
& & -
\frac{\Lambda}{6}\epsilon_{abc}\hat{E}^{a}{}_{i}\hat{E}^{b}{}_{j}\hat{E}^{c}{}_{k}\epsilon^{ijk}=0,
\label{eq615}
\end{eqnarray}
which remains polynomial.

One expects that $\psi_{CS}$ also satisfy the constraints in the
new representation, however one must consider it as function of
$B$, since in its original form it is a function of $A$. However,
we end at this point and leave this problem for a possible next
work.

\section{Conclusions and prospects}

We have shown that it is possible to write the dynamic equations
of general relativity in terms of new variables, which are not
canonical. We obtained a Poisson bracket (associated with the
canonical Hamiltonian structure) and it was shown that it yields
the expected relations between the Hamiltonian and any functional
of the field. The constraint algebra was studied in terms of the
new field variables. The only disadvantage is that we cannot write
explicitly the connection $A$ in terms of $B$ in general and we
are restricted to the cases which it is possible (see $e.g.$ Ref.\
\cite{Freed}).

Usually the ADM formalism is considered as a  metric
representation and the Ashtekar formalism as a connection
representation; the formalism presented here can be considered as
a curvature representation to describe gravity. However, it is
necessary to point out that in this framework we do not have an
action that leads to the new Hamiltonian formulation of gravity.

We have replaced the new variables by operators, however we do not
know at present whether they could help in gaining information
about quantum gravity. This is an interesting problem for a future
work.

Finally, we point out that a similar treatment to that followed
here, is also applicable in Yang-Mills theory \cite{GFRRR}.\\

\section*{Acknowlegments}
I am indebted to Dr.\ G.\ F.\ Torres del Castillo for drawing my
attention on this matter and for his many illuminating remarks and
suggestions on the manuscript. I would also like to thank Dr.\ R.\
Cartas and Dr.\ A.\ Rosado for constant interest in my work. I
acknowledge financial support provided by CONACyT.


\begin{thebibliography}{9}
\bibitem{Ash1} A.\ Ashtekar, {\it Phys. Rev. Lett.} {\bf 57} (1986)
2244.

\bibitem{Ash2} A.\ Ashtekar, {\it Phys.\ Rev.} D (1987) {\bf 36}
1587.

\bibitem{ADM} R.\ Arnowitt, S.\ Deser and C.\ Misner {\it
Gravitation: an introduction to current research}, Ed. L. Witten,
(New York: Wiley, 1962) p.\ 227.

\bibitem{Olv} P.\ J.\ Olver, {\it Applications of Lie Groups to
Differential Equations}, (New York: Springer-Verlag, 1986).

\bibitem{GF1} G.\ F.\ Torres del Castillo, {\it Rev.\ Mex.\ F{\'\i}s.} {\bf
37} (1991) 165.

\bibitem{GF2} G.\ F.\ Torres del Castillo, {\it Rev. Mex. F{\'\i}s.} {\bf
37} (1991) 443.

\bibitem{RRR} R.\ Rosas-Rodr{\'\i}guez, {\it J. Phys.: Conf. Series} {\bf
24} (2005) 231.

\bibitem{MM} M.\ Montesinos, {\it J. Phys.: Conf. Series} {\bf
24} (2005) 44.

\bibitem{MM2} M.\ Mondrag\'on and M.\ Montesinos, {\it Int.\ J.\ Mod.\  Phys.} A, {\bf
19} (2004) 2473.

\bibitem{Wald} R.\ M.\ Wald, {\it General Relativity},
(Chicago: The University of Chicago Press, 1984)

\bibitem{Kuch} K.\ Kuchar, {\it J. Math. Phys.} {\bf 17} (1976) 777, 792,
801; {\it J. Math. Phys.} {\bf 18} (1997) 1589.

\bibitem{Ash3} A.\ Ashtekar, {\it
Lectures on Non Perturbative Canonical Gravity}, (Singapore: World
Scientific, 1991).

\bibitem{Cap} R.\ Capovilla, {\it Proc. IV Mexican Workshop on
Particles and Fields}, (Merida, 1993), p.\ 217.

\bibitem{LMMR} D.\ Lewis, J.\ E.\ Marsden, R.\ Montgomery and T.\
Ratiu, {\it Physica} D {\bf 18} (1986) 391.

\bibitem{Sol} V.\ O.\ Soloviev, {\it J. Math. Phys.} {\bf 34} (1993)
5747.

\bibitem{W-Y} T.\ T.\ Wu and C.\ N.\ Yang, {\it Phys. Rev.} D {\bf 12} (1975)
3845.

\bibitem{Freed} D.\ Z.\ Freedman and R.\ R.\ Khuri, {\it Phys. Lett.} B {\bf 329} (1994)
263.

\bibitem{Gamb} R.\ Gambini and J.\ Pullin, {\it Loops, Knots, Gauge Theories
and Quantum Gravity}, (Cambridge University Press, 2000).

\bibitem{Smo} T.\ Jacobson and L.\ Smolin, {\it Nucl. Phys.} B {\bf 299} (1998)
295.

\bibitem{GFRRR} R.\ Rosas Rodr{\'\i}guez, ``Alternative Hamiltonian Representation for
Yang-Mills Theories" ({\it in preparation}).
\end{thebibliography}
\end{document}